\documentclass{PoS}

\title{A rephasing invariant study of neutrino mixing}

\ShortTitle{A rephasing invariant study of neutrino mixing}

\author{\speaker{S. H. Chiu}\\
        Physics Group, CGE, Chang Gung University, 
Taoyuan 333, Taiwan\\
        E-mail: \email{schiu@mail.cgu.edu.tw}}

\abstract{We derive a set of renormalization group equations (RGE) 
for Dirac neutrinos using a rephasing invariant parametrization.
The symmetric properties of these equations under flavor permutation
facilitate the derivation of some exact and approximate RGE invariants.  
Even though the complete analytical solutions for the RGE are
unavailable, we provide a numerical example that illustrate the evolution of 
the neutrino mixing parameters.}

\FullConference{The European Physical Society Conference on High Energy Physics\\
		22--29 July 2015\\
		Vienna, Austria}

\begin{document}

\section{Introduction}

%It is well motivated that new physics beyond the standard model (SM)
%should apply between the electroweak scale ($\sim 10^{2}$ GeV) and the Planck
%scale ($\sim 10^{2}$ GeV) to account for the unknown origin of the leptonic
%flavor mixing and neutrino masses (see, e.g., Ref.\cite{Ohlsson:2013xva}. 
The bridging over the high energy models
and the measured neutrino parameters at low energies will only be possible if the 
renormalization effects of neutrino parameters are
properly considered. As the energy scale changes, one expects the formulation
of theory to evolve according to the renormalization group equations (RGE).
Intensive effort has thus been devoted to the study of the RGE effect 
of the neutrino parameters 
(see, e.g., \cite{Casas:1999tg,Chankowski:1999xc,Antusch:2003kp,Lindner:2005as,Luo:2005fc,Ray:2010rz}).  %%%%%%%%%%%%%%
The formulation is in general rather complicated under the framework of standard parametrization.
While it may appear that the choice of parametrization is not important since the resultant
physical quantities are independent of the parametrization
and must be expressed as rephasing invariant combinations of the parameters.
However, for instance, when one considers the relations between parameters
such as in the RGE for neutrino mixing which depends on the energy scale,
a particular choice of parametrization may be advantageous over others. 

Whether neutrinos are Dirac or Majorana particles has been one 
of the most important open questions of neutrino physics. 
Both scenarios are well motivated from the theoretical perspectives, 
as can be seen from a wealth of studies in the literature 
(see, e.g., \cite{Mohapatra:2004vr,Smirnov:2004hs}). %%%%%%%%%%%%%%%%%%%%%%%%%
Although many favor the well known see-saw mechanism as a support for Majorana
neutrinos, there exists other models that provide theoretical grounds for
the possibility of Dirac neutrinos (see, e.g., Ref.\cite{Lindner:2005as} and the references therein).
It remains undetermined
unless further experimental evidence is confirmed.
However, investigating the theoretical implication for both scenarios 
will provide indispensable hints at a more fundamental level.

It is the purpose of this work to investigate the RGE evolution of the Dirac neutrinos
using a new set of rephasing invariant parameters. 
It is found that the evolution equations for many of the neutrino parameters take
similar matrix expressions, which may facilitate both the analytical
and the numerical study of the RGE evolution, and may provide an alternative and transparent view
to the underlying physics. We discuss some intriguing properties of the parameters and derive
RGE invariants which are combinations of neutrino masses and mixing.
For illustration purpose, we also show examples of the numerical solutions under specific initial 
conditions.

% Sec II
%%%%%%%%%%%%%%%%%%%%%%%%%%%%%%%%%%%%%%%%%%%%%%%%%%%%%%%%
\section{A rephasing invariant parametrization and its properties}  

The rephasing invariant combinations of elements $V_{ij}$ for the neutrino mixing matrix
$V$ can be constructed by %%%%%%%%%%%%%%%%%Ref
the product \cite{Kuo:05,CKL,CK10,CK11,Chiu:2012uc}
\begin{equation}\label{gamma}
\Gamma_{ijk}=V_{1i}V_{2j}V_{3k}=R_{ijk}-iJ,
\end{equation}
where the common imaginary part is identified with the Jarlskog invariant,
and the real parts are defined as
\begin{equation}
(R_{123},R_{231},R_{312};R_{132},R_{213},R_{321})=(x_{1},x_{2},x_{3};y_{1},y_{2},y_{3}).
\end{equation}
The $(x_{i},y_{j})$ variables are constrained by two conditions:
\begin{equation}\label{con1}
det V=(x_{1}+x_{2}+x_{3})-(y_{1}+y_{2}+y_{3})=1,
\end{equation}
\begin{equation}\label{con2}
(x_{1}x_{2}+x_{2}x_{3}+x_{3}x_{1})-(y_{1}y_{2}+y_{2}y_{3}+y_{3}y_{1})=0,
\end{equation}
and they are related to the Jarlskog invariant \cite{Jar:85},
\begin{equation}
J^{2}=x_{1}x_{2}x_{3}-y_{1}y_{2}y_{3}.
\end{equation}
In addition, the $(x_{i},y_{j})$ variables are bounded by $\pm 1$: $-1 \geq (x_{i},y_{j}) \geq +1$,
with $x_{i} \geq y_{j}$ for any pair of $(i,j)$.  

It is convenient to write $|V_{ij}|^{2}$ in a matrix form with elements $x_{i}-y_{j}$:
\begin{equation}
W=[|V_{\alpha i}|^{2}]=
\left(\begin{array}{ccc}
   x_{1}-y_{1} & x_{2}-y_{2} & x_{3}-y_{3} \\
   x_{3}-y_{2} & x_{1}-y_{3} & x_{2}-y_{1} \\
    x_{2}-y_{3} & x_{3}-y_{1} &x_{1}-y_{2} \\
    \end{array}
    \right)
    \end{equation}
The matrix of the cofactors of $W$, denoted as $w$ with $w^{T}W= (detW)I$, is given by
\begin{equation}
w=[|V_{\alpha i}|^{2}]=
\left(\begin{array}{ccc}
   x_{1}+y_{1} & x_{2}+y_{2} & x_{3}+y_{3} \\
   x_{3}+y_{2} & x_{1}+y_{3} & x_{2}+y_{1} \\
    x_{2}+y_{3} & x_{3}+y_{1} &x_{1}+y_{2} \\
    \end{array}
    \right)
    \end{equation}
The elements of $w$ are also bounded, $-1 \geq w_{\alpha i} \geq +1$.

One may further obtain useful expressions of the rephasing invariant combination formed by
products of four mixing elements, 
\begin{equation}
\pi_{ij}^{\alpha \beta}=V_{\alpha i}V_{\beta j}V_{\alpha j}^{*}V_{\beta i}^{*},
\end{equation}
which can be reduced to 
\begin{eqnarray}
\pi_{ij}^{\alpha \beta} & = & |V_{\alpha i}|^{2}|V_{\beta j}|^{2}-
 \sum_{\gamma k}\epsilon _{\alpha \beta \gamma}\epsilon_{ijk}V_{\alpha i}V_{\beta j}V_{\gamma k} \nonumber \\
   & = & |V_{\alpha j}|^{2}|V_{\beta i}|^{2}+
 \sum_{\gamma k}\epsilon _{\alpha \beta \gamma}\epsilon_{ijk}V_{\alpha j}^{*}V_{\beta i}^{*}V_{\gamma k}^{*},
 \end{eqnarray}
where the second term in either expression is one of the $\Gamma$'s ($\Gamma^{*}$'s) defined in Eq.~(\ref{gamma}).

%%%%%%%%%%%%%%%%%%%%%%%%%%%%%%%%%%%%%%%%%%%%%%%%%%%%%%%%%%%%%%%%%%
%Sec III

\section{RGE for the neutrino parameters} 

The RGE for the Hermitian matrix $M\equiv Y_{\nu}^{\dag}Y_{\nu}$, where $Y_{\nu}$ is the neutrino
Yukawa coupling matrix, is given by %%%%%%%%%%%%%%%%%%%%%%%Ref
\begin{equation}
16\pi^{2}\frac{dM}{dt}=\alpha M+P^{\dag}M+MP
\end{equation} 
at the one-loop level. Here $\alpha$ is real and model-independent, 
$P=C_{\nu}^{e}Y_{l}^{\dag}Y_{l}+C_{\nu}^{\nu}Y_{\nu}^{\dag}Y_{\nu}$, with model-dependent
coefficients $C_{\nu}^{l}$ and $C_{\nu}^{\nu}$. The Yukawa coupling matrices for
neutrinos and charged leptons are denoted as $Y_{\nu}$ and $Y_{l}$, respectively.
One may choose the flavor basis where the charged-lepton Yukawa coupling matrix $Y_{l}$
is diagonal, and 
the matrix $M$ can be diagonalized by the matrix $V$:
\begin{equation}
M=V[diag(h^{2}_{1},h^{2}_{2},h^{2}_{3})]V^{\dag},
\end{equation}
where $h^{2}_{i}$ are the eigenvalues of $M$.
%Note that the chosen flavor basis is stable against the RGE running since
%the evolution equations for the charged leptons can be arranged in
%an equation of diagonal matrices. Hence $Y_{l}Y^{\dag}_{l}$ remains 
%diagonal (if one ignores the tiny $Y_{\nu}Y^{\dag}_{\nu}$),
%while its eigenvalues get modified as the energy scale changes.

The evolution of the mixing matrix $V$ satisfies the relation, 
\begin{equation}\label{eq:T}
dV/dt=VT,
\end{equation}
here the matrix $T$ is anti-Hermitian.
One may define $\mathcal{D}=16\pi^{2}\frac{d}{dt}$ with $t=\ln(\mu/M_{W})$,
and compares the diagonal elements of 
$\mathcal{D}(V[diag(h^{2}_{1},h^{2}_{2},h^{2}_{3})]V^{\dag})$ with 
that of $\mathcal{D}M$ to obtain
\begin{equation}
\mathcal{D}h^{2}_{i}=h_{i}^{2}[\alpha 
+2 C^{l}_{\nu}(|V_{1i}|^{2}f_{1}^{2}+|V_{2i}|^{2}f_{2}^{2}+|V_{3i}|^{2})f_{3}^{2}],
\end{equation}
where $f^{2}_{i}$ are the eigenvalues of the matrix $Y_{l}^{\dag}Y_{l}$,
and the $C^{\nu}_{\nu}$ terms have been ignored.
From the off-diagonal elements, we obtain the expression of $T_{ij}$:
\begin{equation}\label{eq:HP}
T_{ij}=-H_{ij}P'_{ij}/(16\pi^{2}),
\end{equation}
with $P'= V^{\dag}PV$ and
\begin{equation}\label{eq:H}
H_{ij}=\frac{h_{i}^{2}+h_{j}^{2}}{h_{i}^{2}-h_{j}^{2}}.
\end{equation}

%%%%%%%%%%%%%%%%%%%%%%%%%%%%%%%%%%%%%%%%%%%%%%%%%%%%%%Table II
%%%%%%%%%%%%%%%%%%%%%%%%%%%%%%%%%%%%%%%%%%%%%%%%%%%%%%%Table I
 \begin{table}
 \begin{center}
 \begin{tabular}{cc}
$[A_{i}]$ & $[A'_{i}]$    \\     \hline \hline  
\vspace{0.25in}   
  $[A_{1}]=\left(\begin{array}{ccc}
   2x_{1}y_{1} & x_{1}x_{2}+y_{2}y_{3} & x_{1}x_{3}+y_{2}y_{3} \\
    x_{1}x_{3}+y_{1}y_{2} & 2x_{1}y_{3} & x_{1}x_{2}+y_{1}y_{2} \\
    x_{1}x_{2}+y_{1}y_{3} & x_{1}x_{3}+y_{1}y_{3} & 2x_{1}y_{2} \\
    \end{array}
    \right)$, 
 &  $[A'_{1}]=\left(\begin{array}{ccc}
    2x_{1}y_{1} & x_{2}x_{3}+y_{1}y_{2} & x_{2}x_{3}+y_{1}y_{3} \\
    x_{1}x_{3}+y_{1}y_{2} & x_{1}x_{3}+y_{1}y_{3} & 2x_{2}y_{1} \\
    x_{1}x_{2}+y_{1}y_{3} & 2x_{3}y_{1} & x_{1}x_{2}+y_{1}y_{2} \\
    \end{array}
    \right)$

 \\ \vspace{.25in} 
   $[A_{2}]=\left(\begin{array}{ccc}
    x_{1}x_{2}+y_{1}y_{3}&2x_{2}y_{2} & x_{2}x_{3}+y_{1}y_{3} \\
    x_{2}x_{3}+y_{2}y_{3} & x_{1}x_{2}+y_{2}y_{3} & 2x_{2}y_{1} \\
    2x_{2}y_{3} & x_{2}x_{3}+y_{1}y_{2} & x_{1}x_{2}+y_{1}y_{2} \\
    \end{array}
    \right)$,   
  &  $[A'_{2}]=\left(\begin{array}{ccc}
 x_{1}x_{3}+y_{1}y_{2}  & 2x_{2}y_{2} & x_{1}x_{3}+y_{2}y_{3} \\
   2x_{3}y_{2} & x_{1}x_{2}+y_{2}y_{3} & x_{1}x_{2}+y_{1}y_{2} \\
    x_{2}x_{3}+y_{2}y_{3} & x_{2}x_{3}+y_{1}y_{2} & 2x_{1}y_{2} \\
    \end{array}
    \right)$

 \\ \vspace{.25in}
       
    $[A_{3}]=\left(\begin{array}{ccc}
    x_{1}x_{3}+y_{1}y_{2} & x_{2}x_{3}+y_{1}y_{2} & 2x_{3}y_{3} \\
    2x_{3}y_{2} & x_{1}x_{3}+y_{1}y_{3} & x_{2}x_{3}+y_{1}y_{3} \\
    x_{2}x_{3}+y_{2}y_{3} & 2x_{3}y_{1} & x_{1}x_{3}+y_{2}y_{3} \\
    \end{array}
    \right)$,
 &   $[A'_{3}]=\left(\begin{array}{ccc}
    x_{1}x_{2}+y_{1}y_{3} & x_{1}x_{2}+y_{2}y_{3} & 2x_{3}y_{3} \\
    x_{2}x_{3}+y_{2}y_{3} & 2x_{1}y_{3} & x_{2}x_{3}+y_{1}y_{3} \\
    2x_{2}y_{3} & x_{1}x_{3}+y_{1}y_{3} & x_{1}x_{3}+y_{2}y_{3} \\
    \end{array}
    \right)$  
  
   \\  \hline \hline
   
    \end{tabular}
 \caption{The explicit expressions of the matrices $[A_{i}]$ and $[A'_{i}]$.} 
   \end{center}
 \label{tab1}
   \end{table}

%%%%%%%%%%%%%%%%%%%%%%%%%%%%%%%

%

%%%%%%%%%%%%%%%%%%%%%%%%%%%%%%%

To derive the RGE for neutrino mixing parameters, we may use
Eq.~(\ref{eq:T}), Eq.~(\ref{eq:HP}), and Eq.~(\ref{eq:H}) to obtain
\begin{equation}
\mathcal{D}\Gamma_{ijk}=-[(\sum_{l\neq i}V_{1l}H_{li}P'_{li})V_{2j}V_{3k}+
V_{1i}(\sum_{l\neq j}V_{2l}H_{lj}P'_{lj})V_{3k}+V_{1i}V_{2j}(\sum_{l\neq k}V_{3l}H_{lk}P'_{lk})].
\end{equation}
The real part of $\mathcal{D}\Gamma_{ijk}$ gives rise to $\mathcal{D}x_{i}$ and 
$\mathcal{D}y_{i}$ in the following matrix forms:
\begin{equation}
\mathcal{D}x_{i}=-2C_{\nu}^{l}[\Delta f^{2}_{23},\Delta f^{2}_{31},\Delta f^{2}_{12}][A_{i}][H_{23},H_{31},H_{12}]^{T},
\end{equation}
\begin{equation}
\mathcal{D}y_{i}=-2C_{\nu}^{l}[\Delta f^{2}_{23},\Delta f^{2}_{31},\Delta f^{2}_{12}][A_{i}'][H_{23},H_{31},H_{12}]^{T},
\end{equation}
here $\Delta f^{2}_{ij}=f^{2}_{i}-f^{2}_{j}$, and 
the matrices $[A_{i}]$ and $[A'_{i}]$ are given by Table I.
In addition, the imaginary part leads to
\begin{equation}
\mathcal{D}J^{2}=-2C_{\nu}^{l}[\Delta f^{2}_{23},\Delta f^{2}_{31},\Delta f^{2}_{12}][w][H_{23},H_{31},H_{12}]^{T}
\end{equation}

The evolution equations for $W_{ij}=|V_{ij}|^{2}$ can be derived directly
from $\mathcal{D}x_{i}$ and $\mathcal{D}y_{i}$:
\begin{equation}
\mathcal{D}W_{ij}=
-2C_{\nu}^{l}[\Delta f^{2}_{23},\Delta f^{2}_{31},\Delta f^{2}_{12}][S_{ij}][H_{23},H_{31},H_{12}]^{T},
\end{equation}
where the matrices $[S_{ij}]$ are obtained by using suitable combinations
of $[A_{i}]$ and $[A'_{i}]$. For instance, $[S_{23}]=[A_{2}]-[A'_{1}]$, etc.
In addition, we denote the real part of $\pi_{\gamma k}$ as $\Lambda_{\gamma k}$,
\begin{equation}
\pi^{\alpha \beta}_{ij}\equiv \pi_{\gamma k}=\Lambda_{\gamma k}+iJ.
\end{equation}
Since $Re(\pi^{\alpha \beta}_{ij})$ takes the following forms,
\begin{equation}
Re(\pi^{\alpha \beta}_{ij})=|V_{\alpha i}|^{2}|V_{\beta j}|^{2}-x_{a}=|V_{\beta i}|^{2}|V_{\alpha j}|^{2}+y_{b},
\end{equation}
the average of the two different forms is thus equal to $\Lambda_{\gamma k}$,
\begin{equation}\label{lambda}
\Lambda_{\gamma k}=
\frac{1}{2}(|V_{\alpha i}|^{2}|V_{\beta j}|^{2}+|V_{\alpha j}|^{2}|V_{\beta i}|^{2}-|V_{\gamma k}|^{2}),
\end{equation}
where $(\alpha,\beta, \gamma)$ and $(i,j,k)$ are cyclic permutations.
Furthermore, we may write down the elements of the matrix $[w]$ as
\begin{equation}
w_{\gamma k}=|V_{\alpha i}|^{2}|V_{\beta j}|^{2}-|V_{\alpha j}|^{2}|V_{\beta i}|^{2}.
\end{equation}
The evolution equation for the elements of $[w]$ can then be shown to take the form:
\begin{equation}
\mathcal{D}w_{ij}=-2C_{\nu}^{l}[\Delta f^{2}_{23},\Delta f^{2}_{31},\Delta f^{2}_{12}][G_{ij}][H_{23},H_{31},H_{12}]^{T}
\end{equation}
where the matrices $[G_{ij}]$ are obtained by using suitable combinations
of $[A_{i}]$ and $[A'_{i}]$. For instance, $[G_{21}]=[A_{3}]-[A'_{2}]$, etc. 

%%%%%%%%%%%%%%%%%%%%%%%%%%%%%%%%%%%%%%%%%%%%%%%%%%%%%%%%%%%%%%%%%%%%%%%%%%%%%%%%%%%%%

%%%%%%%%%%%%%%%%%%%%%%%%%%%%%%%%%%%%%%%%%%%%%%%%%%%%%%Table IV

%%%%%%%%%%%%%%%%%%%%%%%%%%%%%%%

%%%%%%%%%%%%%%%%%%%%%%%%%%%%%%%%%%%%%%%%%%%%%%%%%%%%%%%%

\section{The RGE invariants}

The study of RGE invariants has drawn certain attention in recent literature 
(see, e.g., \cite{Demir:2004aq,Harrison:2010mt,Feldmann:2015nia}).   %%%%%%%%%%%%%%%%
The invariants are formed by combinations of physical observables
which remain the same under the RGE running of energy scale.
To search for neutrino RGE invariants using our parametrization, we first define the neutrino mass ratio
$r_{ij}=m_{i}/m_{j}$, where $h_{i}=\sqrt{2}m_{i}/v$ in the SM and $h_{i}=\sqrt{2}m_{i}/(v\cos\beta)$
in the MSSM, with $v\simeq 246$ GeV and $\tan\beta$ is given by the ratio of the two Higgs VEVs 
in MSSM. By calculating $\mathcal{D}(\sinh^{2}\ln r_{ij})$ and 
$\mathcal{D}[\ln(\sinh^{2}\ln r_{ij})]$, we obtain
\begin{equation}
\Sigma_{ij}\mathcal{D}[\ln(\sinh^{2}\ln r_{ij})]=
2C_{\nu}^{l}[\Delta f_{23},\Delta f_{31}, \Delta f_{12}][w][H_{23},H_{31},H_{12}]^{T}.
\end{equation}
Combining this result with the expression of $\mathcal{D}J^{2}$, we find that
\begin{equation}\label{inv-1}
\mathcal{D}[\ln(J^{2} \cdot \sinh^{2}\ln r_{12}\cdot \sinh^{2}\ln r_{23} \cdot \sinh^{2}\ln r_{31})]=0,
\end{equation}
i.e., $J^{2}(\Pi_{ij}\sinh^{2}\ln r_{ij})$
is a RGE invariant.

Furthermore, one notes that any quantity that is identical to $J^{2}$ also gives 
rise to an invariant when
multiplied by $\Pi_{ij}\sinh^{2}\ln r_{ij}$ in Eq.~(\ref{inv-1}).
With the expression of $\Lambda_{\gamma k}$ in Eq.~(\ref{lambda}),
it is straightforward to write down nine different forms of $J^{2}=\pi^{2}_{\gamma k}-\Lambda^{2}_{\gamma k}$,
which correspond to nine different combinations of $(\gamma,k)$,
\begin{eqnarray}
J^{2}& = & \pi^{2}_{\gamma k}-\Lambda^{2}_{\gamma k} \nonumber \\
   & = &  |V_{\alpha i}|^{2}|V_{\beta j}|^{2}|V_{\alpha j}|^{2}|V_{\beta i}|^{2}-\Lambda^{2}_{\gamma k}.
\end{eqnarray}
This leads to nine RGE invariants which consist of $|V_{ij}|^{2}$ and the mass ratios $\ln r_{ij}^{2}$:
\begin{equation}\label{inv}
[|V_{\alpha i}|^{2}|V_{\beta j}|^{2}|V_{\alpha j}|^{2}|V_{\beta i}|^{2}-\Lambda^{2}_{\gamma k}]
(\sinh\ln r_{21}^{2})(\sinh\ln r_{32}^{2})
(\sinh\ln r_{13}^{2})=constant
\end{equation}

One may also find approximate invariants under certain limiting cases.
If $f^{2}_{\alpha} \gg f^{2}_{\beta}, f^{2}_{\gamma}$, 
the evolution equations for $\mathcal{D}[\ln (J^{2})]$ 
and $\mathcal{D}|V_{ij}|^{2}$ lead directly to
\begin{equation}
J^{2}/|V_{\alpha 1}|^{2}|V_{\alpha 2}|^{2}|V_{\alpha 3}|^{2}\simeq constant.
\end{equation}
Furthermore, if the neutrino masses satisfy the hierarchical condition: 
$H_{ij} \gg H_{jk},H_{ki}$, then $(d \ln J^{2}/dt) + (d \ln(\sinh^{2}\ln r_{ij})/dt)\approx 0$
and the approximate invariant follows:
\begin{equation}
 J^{2}(\sinh^{2}\ln r_{ij})=constant.
 \end{equation}

%Note that the possible RGE effects that may distinguish the normal or inverted mass ordering 
%become unobservable in this case since $H_{12}$ maintains the same sign for both mass orderings.

%%%%%%%%%%%%%%%%%%%%%%%%%%%%%%%%%%%%%%%%%%%%%%%%%%%%%%%%%%%%%%%%%%%%%%%%%%%%%%%%%%%%%
%Sec IV

\begin{figure}
\includegraphics[width=.9\textwidth]{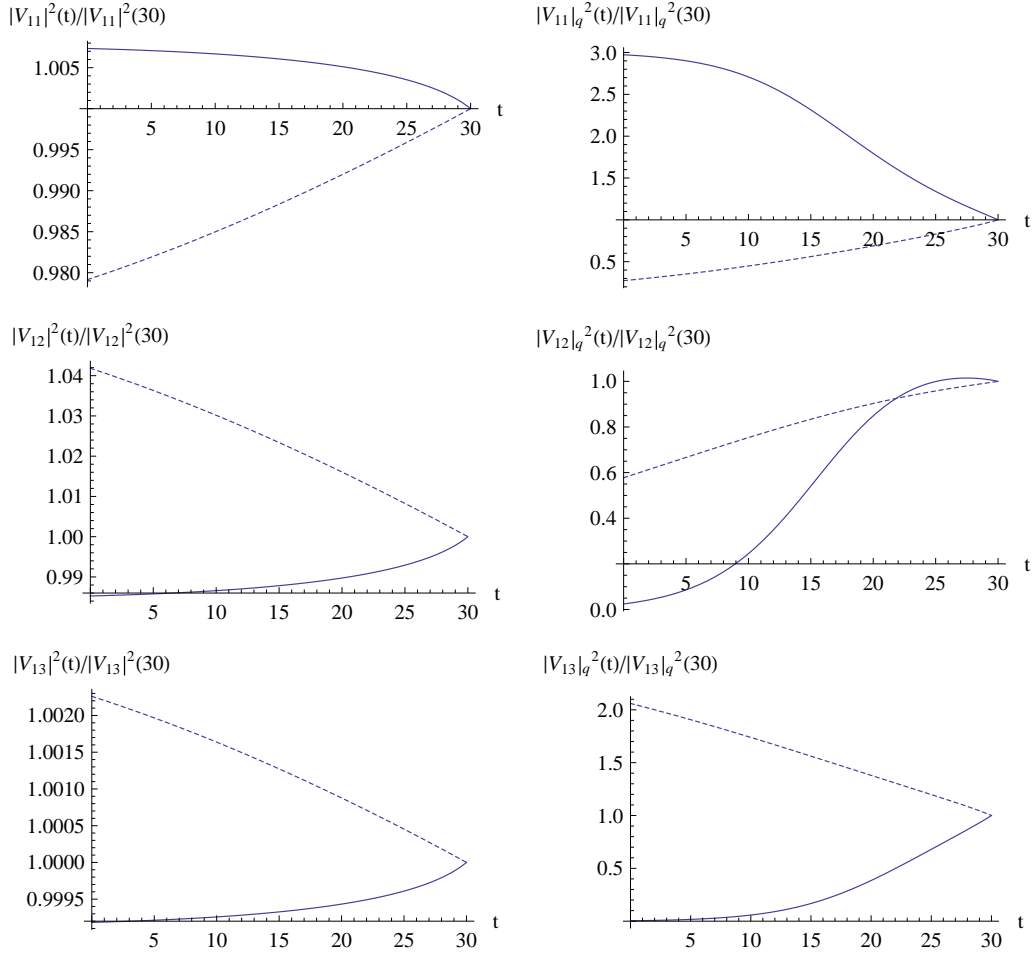}
\caption{The evolution of $|V_{1i}|^{2}(t)/|V_{1i}|^{2}(t=30)$ for neutrinos
and $|V_{1i}|^{2}_{q}(t)/|V_{1i}|^{2}_{q}(t=30)$ for quarks under SM (solid) and MSSM (dashed).}
\label{fig1}
\end{figure}

\section{Numerical examples}

For the numerical calculation of this work,
the quark RGEs using the $(x,y)$ parametrization shall be adopted from Ref.\cite{Chiu:2008ye}.
To assess the general nature of the RGE, 
it seems more appropriate to start from a point with fast evolution, 
so that most changes may be accomplished in its neighborhood,
with minor corrections afterwards.
It is pointed out that the low energy physics values are close to a fixed point
of the RGE, and it is crucial to study how they are approached from the high energy values.   
%However, without detailed knowledge of the initial values at high energy in any 
%theoretical framework, it is difficult to assign a suitable region 
%of the initial parameter space that yields all the measured values
%as the RGEs evolve down to the low energy. 

For the purpose of illustration, 
we adopt the following initial input parameters at high energy for a numerical example.
(i) The quark masses at $t=30$ are assumed to be near degenerate with $m_{q}\sim 173$ GeV.
(ii) The mixing parameters for quarks are taken to be 
$[x_{1},x_{2};y_{1},y_{2}]=[(1/6)+\epsilon,(1/6)-\epsilon;(-1/6)+\epsilon,(-1/6)+\epsilon]$,
with $\epsilon=0.01$.
(iii) The neutrino masses at $t=30$ are also taken to be near degenerate,
 $m'_{3} \sim m'_{2} \sim m'_{1} \sim 0.05$ eV.
(iv) As for the neutrino mixing parameters, it is found that $(x_{i},y_{j})$ for neutrinos
only evolve slightly from their respective initial values.
We therefore adopt $[x_{1},x_{2};y_{1},y_{2}]=[(1/3)-\epsilon,(1/6)-\epsilon;(-1/3)+\epsilon,(-1/6)+\epsilon]$
as the input at $t=30$ so as to yield reasonable mixing parameters at low energy.

We show in Figure 1 a numerical example that illustrate the evolution 
of $|V_{1i}|^{2}(t)$ for both quarks and neutrinos in the unit of 
the respective initial value at $t=30$: $|V_{1i}|^{2}(t)(30)$.
It is seen that the quark mixing can evolve quite significantly from high to low $t$.
However, since
the neutrino mixing parameters ($x_{i}$ and $y_{i}$) only evolve slightly,
their elements $|V_{ij}|^{2}$ do not evolve much from the initial values at high energy.  
This general behavior is in agreement with the expectation of 
recent study (see, e.g., Ref\cite{Ohlsson:2013xva}).  %%%%%%%%%%%%%%%%%%%%%%%%%
Despite the tiny evolution, the SM and MSSM models suggest opposite trends
of evolution for $|V_{ij}|^{2}$ due to the opposite signs of $C$ in each model. 
It is seen that $|V_{ij}|^{2}$ decreases under one model while increases under
the other.

 %%%%%%%%%%%%%%%%%%%%%%%%%%%%%%%%%%%%%%%%%%%%%%%%%%%%%%%%%%%%%%%%%%%%%%

\section{conclusion}

We have been able to derive and analyze the RGEs for Dirac neutrinos using a newly introduced
set of parameters.  Specifically, we show the analytical expressions for 
$\mathcal{D}h^{2}_{i}$ (or $\mathcal{D}m^{2}_{i}$), $\mathcal{D}x_{i}$, $\mathcal{D}y_{i}$,
$\mathcal{D}J^{2}$, $\mathcal{D}|V_{ij}|^{2}$ (or $\mathcal{D}W_{ij}$), and $\mathcal{D}w_{ij}$
in a matrix form.
Certain intriguing properties become transparent
under this parametrization framework.
The matrix equations are shown to be highly symmetric among
various parameters, and certain intriguing properties become transparent
under this parametrization framework. This greatly
facilitates both the theoretical and numerical study of the 
RGE effects for the neutrinos.

Measurable physical quantities are independent of the framework of parametrization,
and one should be able to formulate these observales in an invariant form. 
Based on the RGE of the parameters, we derive several RGE invariants which
are combinations of the mass ratios, $r_{ij}$, and the mixing elements of neutrinos, $|V_{ij}|^{2}$.
In particular, such invariants are presented by Eq.~(\ref{inv-1}) and Eq.~(\ref{inv}).
Approximate invariants are also presented.
We also provide a numerical example that
illustrate the RGE effects of the mixing elements for quarks and neutrinos.
It should be pointed out that ceratin RGE evolution may be sensitive to the input parameters
and that fine-tuning may be required in order to obtain reasonable values at
low energy for all the parameters. 
%This difficulty may be ascribed to
%the situation that a comprehensive and satisfactory model at high energies is still unavailable.
%In addition, the renormalization group evolution may not
%necessarily be the only explanation to the observed parameters at low energy.

This rephasing invariant parametrization may be further applied to the formulation
of RGEs for the Majorana neutrinos, which requires two more phases (or parameters).
We shall present the details of related work in a future study.

\acknowledgments                 
SHC would like to thank Prof. T. K. Kuo for inspirational collaboration. 
This work is supported by the Ministry of Science and Technology of Taiwan, 
Grant No.: MOST 104-2112-M-182-004.

\end{document}